\documentstyle[12pt,epsf]{article}
 \newcommand{\insertplot}[5]{\begin{figure}
 \hfill\hbox to 0.05in{\vbox to #5in{\vfill
 \inputplot{#1}{#4}{#5}}\hfill}
 \hfill\vspace{-.1in}
 \caption{#2}\label{#3}
 \end{figure}}
 \newcommand{\inputplot}[3]{
 \special{ps: plotfile #1}
}

\begin{document}

\title{SU(3) Einstein-Yang-Mills Sphalerons and Black Holes}
\vspace{1.5truecm}
\author{
{\bf Burkhard Kleihaus$^1$, Jutta Kunz$^{1,2}$ and Abha Sood$^1$}\\
$^1$Fachbereich Physik, Universit\"at Oldenburg, Postfach 2503\\
D-26111 Oldenburg, Germany\\
$^2$
Instituut voor Theoretische Fysica, Rijksuniversiteit te Utrecht\\
NL-3508 TA Utrecht, The Netherlands}


\maketitle
\vspace{1.0truecm}

\begin{abstract}
In the SU(3) Einstein-Yang-Mills system sequences of
static spherically symmetric regular solutions
and black hole solutions exist for both the SU(2) and the SO(3) embedding.
We construct the lowest regular solutions of the SO(3) embedding,
missed previously, and the corresponding black holes.
The SO(3) solutions are classified according to
their boundary conditions
and the number of nodes of the matter functions.
Both, the regular and the black hole solutions
are unstable.
\end{abstract}
\vfill
\noindent {Utrecht-Preprint THU-95/8} \hfill\break
\vfill\eject

\section{Introduction}

The SU(2) Einstein-Yang-Mills system
possesses a sequence of static spherically symmetric
regular particle-like solutions \cite{bm},
which are unstable \cite{strau1,volkov1}.
The $n$-th solution of the sequence has $n$ nodes and
$2n$ unstable modes \cite{lav,volkov4}.
The lowest solution has been interpreted
in analogy to the electroweak sphaleron \cite{km}
as the top of a barrier between vacua \cite{volkov1}.

Beside the regular solutions
the SU(2) Einstein-Yang-Mills system possesses
static spherically symmetric black hole solutions.
There are in fact two different types of black hole solutions
having the same mass.
These are the Schwarzschild black holes
with vanishing gauge fields
and the SU(2) coloured black holes \cite{volkov,bizon1,kuenzle1}.
Like the regular solutions
the coloured black hole solutions are unstable \cite{strau2,volkov5,volkov4}.
Thus for a certain range of masses the system possesses
two distinct types of black holes,
providing a counterexample to the ``no-hair conjecture''
for black holes, unless the coloured black holes
are discarded as a counterexample because of their instability.

Here we consider static spherically symmetric
regular solutions and black holes
of the SU(3) Einstein-Yang-Mills system
(with vanishing time component of the gauge field).
Such solutions are obtained for both,
the SU(2) embedding and the SO(3) embedding.
The SU(2) embedding reproduces the known SU(2) solutions,
while the SO(3) embedding leads to new interesting solutions.
Missing the lowest regular solutions,
several regular SO(3) solutions have been found previously
by K\"unzle \cite{kuenzle}, but he did not succeed in obtaining
the corresponding black holes.

We here construct the first few solutions
of a new class of regular SO(3) solutions.
These solutions include the lowest regular SO(3) solution,
which we have obtained first as a limiting solution
of the SU(3) Einstein-Skyrme system \cite{kks}
(in analogy to the SU(2) case \cite{bizon}).
We classify the SO(3) solutions according to
their boundary conditions
and the number of nodes of the matter functions.

Analogous to the SU(2) Einstein-Yang-Mills system
asymptotically flat SO(3) black hole solutions
emerge from the regular solutions by requiring regularity
at a finite event horizon.
We construct the black hole solutions corresponding
to the first few solutions with the least nodes.

The stability of regular and black hole solutions
of arbitrary gauge groups has been studied
recently \cite{strau3}.
We apply the theorems of Ref.~\cite{strau3}
to demonstrate the instability of both
regular and black hole SO(3) solutions.

\section{SU(3) Einstein-Yang-Mills Equations of Motion}

We consider the SU(3) Einstein-Yang-Mills action
\begin{equation}
S=S_G+S_M=\int L_G \sqrt{-g} d^4x + \int L_M \sqrt{-g} d^4x
\   \end{equation}
with
\begin{equation}
L_G=\frac{1}{16\pi G}R
\ , \end{equation}
and the matter Lagrangian
\begin{equation}
L_M= -\frac{1}{2} {\rm Tr} (F_{\mu\nu} F^{\mu\nu})
\ , \end{equation}
where
\begin{equation}
F_{\mu\nu}= \partial_\mu A_\nu - \partial_\nu A_\mu
            - i e [A_\mu,A_\nu]
\ , \end{equation}
\begin{equation}
A_\mu = \frac{1}{2} \lambda^a A_\mu^a
\ , \end{equation}
and $e$ is the coupling constant.
Variation of the action eq.~(1) with respect to the metric
$g_{\mu\nu}$ and the gauge field $A_\mu$
leads to the Einstein equations and the matter field equations.

To construct static spherically symmetric
regular solutions and black holes
we employ Schwarz\-schild-like coordinates and adopt
the spherically symmetric metric
\begin{equation}
ds^2=g_{\mu\nu}dx^\mu dx^\nu=
  -A^2N dt^2 + N^{-1} dr^2 + r^2 (d\theta^2 + \sin^2\theta d\phi^2)
\ , \end{equation}
with
\begin{equation}
N=1-\frac{2m}{r}
\ . \end{equation}
Generalized spherical symmetry for the gauge field is realized
by embedding the SU(2) or the SO(3) generators $T_i$ in SU(3).
In the SU(2)-embedding
$\vec T = \frac{1}{2} (\lambda_1, \lambda_2, \lambda_3)$,
and the ansatz for the gauge field
has vanishing time component \cite{bm},
\begin{eqnarray}
A_0 &=& 0
\ , \nonumber\\
A_i &=& \frac{1-w(r)}{2re} (\vec e_r \times \vec \tau)_i
\ , \label{su2} \end{eqnarray}
with the SU(2) Pauli matrices
$\vec \tau = (\tau_1,\tau_2,\tau_3)$.
In the SO(3)-embedding
$\vec T = (\lambda_7, -\lambda_5, \lambda_2)$,
and the corresponding ansatz for the gauge field
with vanishing time component is
\begin{eqnarray}
A_0 &=& 0
\ , \nonumber\\
A_i &=& \frac{2-K(r)}{2re} (\vec e_r \times \vec \Lambda)_i
       +\frac{H(r)}{2re} \Bigl[ (\vec e_r \times \vec \Lambda)_i,
       \vec e_r \cdot \vec \Lambda) \Bigr]_+
\ , \label{so3} \end{eqnarray}
where $[\ , \ ]_+$ denotes the anticommutator, and
$\vec \Lambda = (\lambda_7,-\lambda_5,\lambda_2)$.

The SU(2)-embedding, eq.~(\ref{su2}),
leads to the well studied SU(2) Einstein-Yang-Mills equations
\cite{bm,volkov,bizon1,kuenzle1}.
To obtain the SU(3) Einstein-Yang-Mills equations
for the SO(3)-embed\-ding, eq.~(\ref{so3}),
we also employ the $tt$ and $rr$
components of the Einstein equations,
yielding for the metric functions
\begin{equation}
\mu'=  N (K'^2 +H'^2)
  + \frac{1}{8 x^2} \left(
     \left(K^2+H^2-4 \right)^2 +12 K^2 H^2 \right)
\ , \end{equation}
\begin{equation}
 A' =  \frac{2}{x} (K'^2 +H'^2) A
\ , \label{eqa} \end{equation}
where we have introduced the dimensionless mass function
\begin{equation}
\mu=\frac{e}{\sqrt{4\pi G}} m = \frac{e m_{\rm Pl}}{\sqrt{4\pi}} m
\   \end{equation}
and the dimensionless coordinate
$x=(e/\sqrt{4\pi G}) r$,
and the prime indicates the derivative
with respect to $x$.
For the matter field functions we obtain the equations
\begin{equation}
(ANK')' = \frac{1}{4 x^2} A K \left( K^2+7H^2-4 \right)
\ , \end{equation}
\begin{equation}
(ANH')' = \frac{1}{4 x^2} A H \left( H^2+7K^2-4 \right)
\ . \end{equation}
With help of eq.~(\ref{eqa}) the metric function $A$
can be eliminated from the matter field equations.
Note, that the equations are symmetric with respect to
an interchange of the functions $K(x)$ and $H(x)$,
and to the transformations $K(x) \rightarrow -K(x)$,
and $H(x) \rightarrow -H(x)$,
yielding degenerate solutions.

Comparing the equations of the SO(3) embedding
to those of the SU(2) embedding \cite{bm}
shows that to each SU(2) solution
there corresponds a scaled SO(3) solution.
Defining
$x = 2 \tilde x$, and $\mu = 2 \tilde \mu$
the functions
$K(x) = 2 w(\tilde x)$, $H(x)=0$
satisfy the SO(3) equations with coordinate $x$,
when the function $w$ satisfies the SU(2) equations
with coordinate $\tilde x$. Thus these SO(3) solutions
have precisely double the mass of their SU(2) counterparts.

\section{Regular Solutions}

Let us first consider the regular solutions
of the SU(3) Einstein-Yang-Mills system.
Requiring asymptotically flat solutions implies
that the metric functions $A$ and $\mu$ both
approach a constant at infinity,
and that the matter functions
approach a vacuum configuration of the gauge field.
We here adopt
\begin{equation}
A(\infty)=1
\ , \end{equation}
thus fixing the time coordinate, and
\begin{equation}
K(\infty)=\pm 2 \ , \quad H(\infty)=0
\ , \label{bc1} \end{equation}
\begin{equation}
K(\infty)=0 \ , \quad H(\infty)=\pm 2
\ . \label{bc2} \end{equation}
At the origin regularity of the solutions requires
\begin{equation}
\mu(0)=0
\ , \end{equation}
and the gauge field functions must satisfy
\begin{equation}
K(0)=\pm 2 \ , \quad H(0)=0
\ , \label{bc3} \end{equation}
\begin{equation}
K(0)=0 \ , \quad H(0)=\pm 2
\ . \label{bc4} \end{equation}
Because of the symmetries of the SO(3) Einstein-Yang-Mills
equations it is sufficient to study solutions
with $K(0)=2$ and $H(0)=0$.
The other boundary conditions lead
to degenerate solutions.

In the following we present some numerical results
for the regular solutions of the SO(3) embedding.
In Table~1 we show the mass $\tilde \mu=\mu /2$
of the lowest SO(3) solutions.
Their ADM mass is
\begin{equation}
m_{\rm ADM}= \mu(\infty) \sqrt{4 \pi} \frac{m_{\rm Pl}}{e}
\ . \end{equation}
We observe, that the two lowest solutions,
missed in the previous analysis by K\"unzle \cite{kuenzle},
have a smaller mass than the lowest scaled SU(2) solution.

To compare with the SO(3) solutions found by K\"unzle
\cite{kuenzle} we note, that his functions $u_1$ and $u_2$
are related to the functions $K$ and $H$ as follows
\begin{equation}
u_1(x)= \frac{K(x)+H(x)}{2}
\ , \end{equation}
\begin{equation}
u_2(x)= \frac{K(x)-H(x)}{2}
\ , \end{equation}
with the boundary conditions at the origin $u_1(0)=u_2(0)=1$,
and at infinity $u_1(\infty)=\pm 1$ and
$u_2(\infty)=\pm 1$.

Let us adopt the classification of
the solutions with respect to their boundary conditions
at infinity,
the nodes $(n_1,n_2)$ of the functions $(u_1,u_2)$
\cite{kuenzle}, and the total number of nodes $n=n_1+n_2$.
We see in Table~1, that the lowest SO(3) solution
has the node structure $(0,1)$, i.~e.~$n=1$.
In contrast, the lowest scaled SU(2) solution,
being the lowest SO(3) solution found by Kuenzle \cite{kuenzle},
has the node structure $(1,1)$, i.~e.~$n=2$.
Naturally, the mass of the SO(3) solution
with one node only is lower than the mass of the
scaled SU(2) solution with $n=2$.
But there is a second SO(3) solution with a lower mass.
This solution has the node structure $(0,2)$,
i.~e.~a total of two nodes like the lowest scaled SU(2) solution.
Evidently, the whole class of solutions
with node structure $(0,n)$
has been missed before \cite{kuenzle}.
This class contains the lowest SO(3) solution,
and for a given total number of nodes,
these new solutions appear to be lowest.

Table~1 further gives the coefficients $\beta_1$ and $\beta_2$
for the numerical integration with the shooting method \cite{kuenzle}
\begin{equation}
u_1(\tilde x)=1+\beta_1 \tilde x^2+\beta_2 \tilde x^3+...
\ , \end{equation}
\begin{equation}
u_2(\tilde x)=1+\beta_1 \tilde x^2-\beta_2 \tilde x^3+...
\ . \end{equation}

\begin{figure}
\centering
\epsfysize=11cm
\mbox{\epsffile{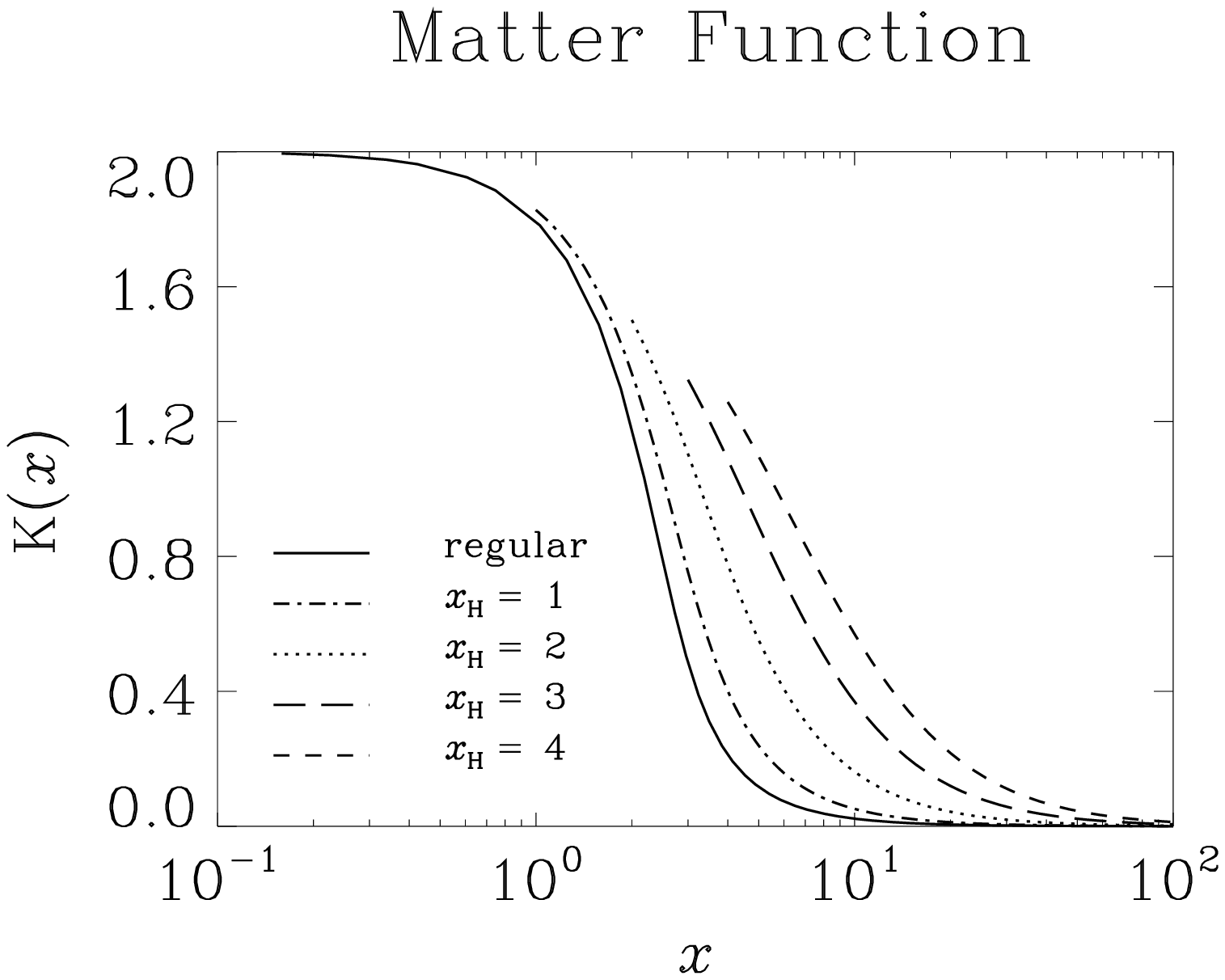}}
\caption{\label{k}
The function $K(x)$
is shown for the regular solution and for the
black hole solutions with horizons
$x_{\rm H}=1$, 2, 3 and 4
as a function of $x$.}
\end{figure}

\begin{figure}
\centering
\epsfysize=11cm
\mbox{\epsffile{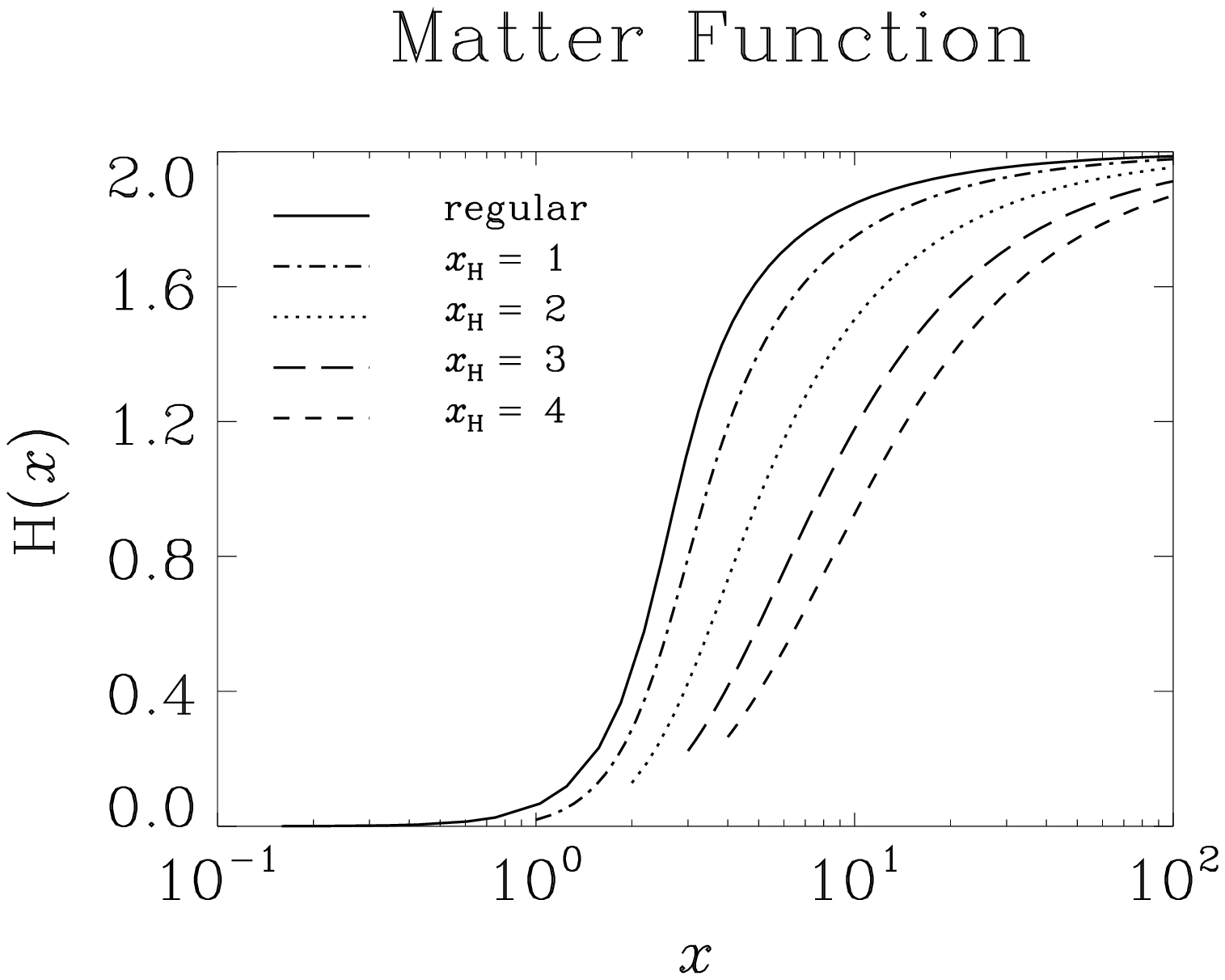}}
\caption{\label{h}
The function $H(x)$
is shown for the regular solution and for the
black hole solutions with horizons
$x_{\rm H}=1$, 2, 3 and 4
as a function of $x$.}
\end{figure}

\begin{figure}
\centering
\epsfysize=11cm
\mbox{\epsffile{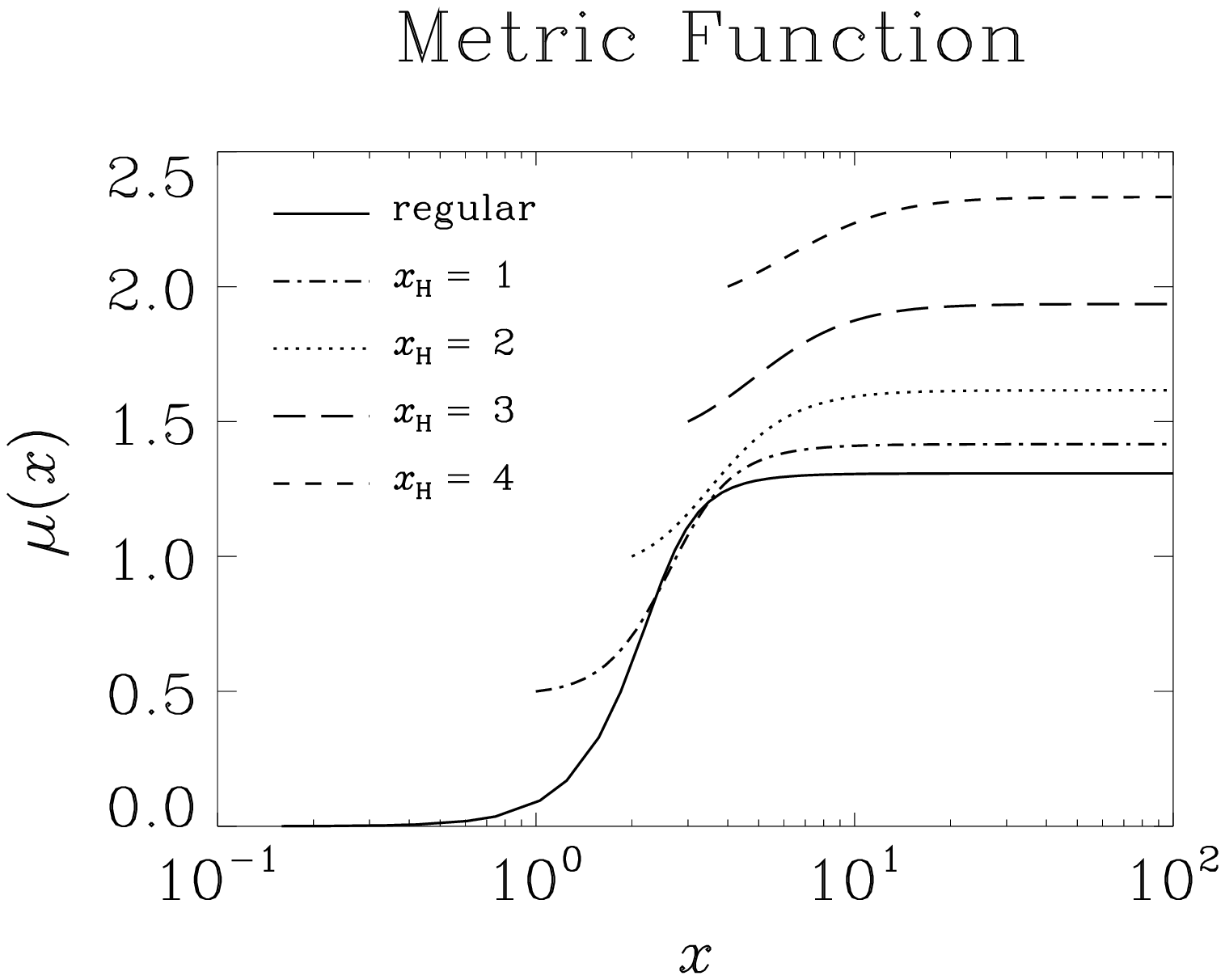}}
\caption{\label{mu}
The function $\mu(x)$
is shown for the regular solution and for the
black hole solutions with horizons
$x_{\rm H}=1$, 2, 3 and 4
as a function of $x$.}
\end{figure}

\begin{figure}
\centering
\epsfysize=11cm
\mbox{\epsffile{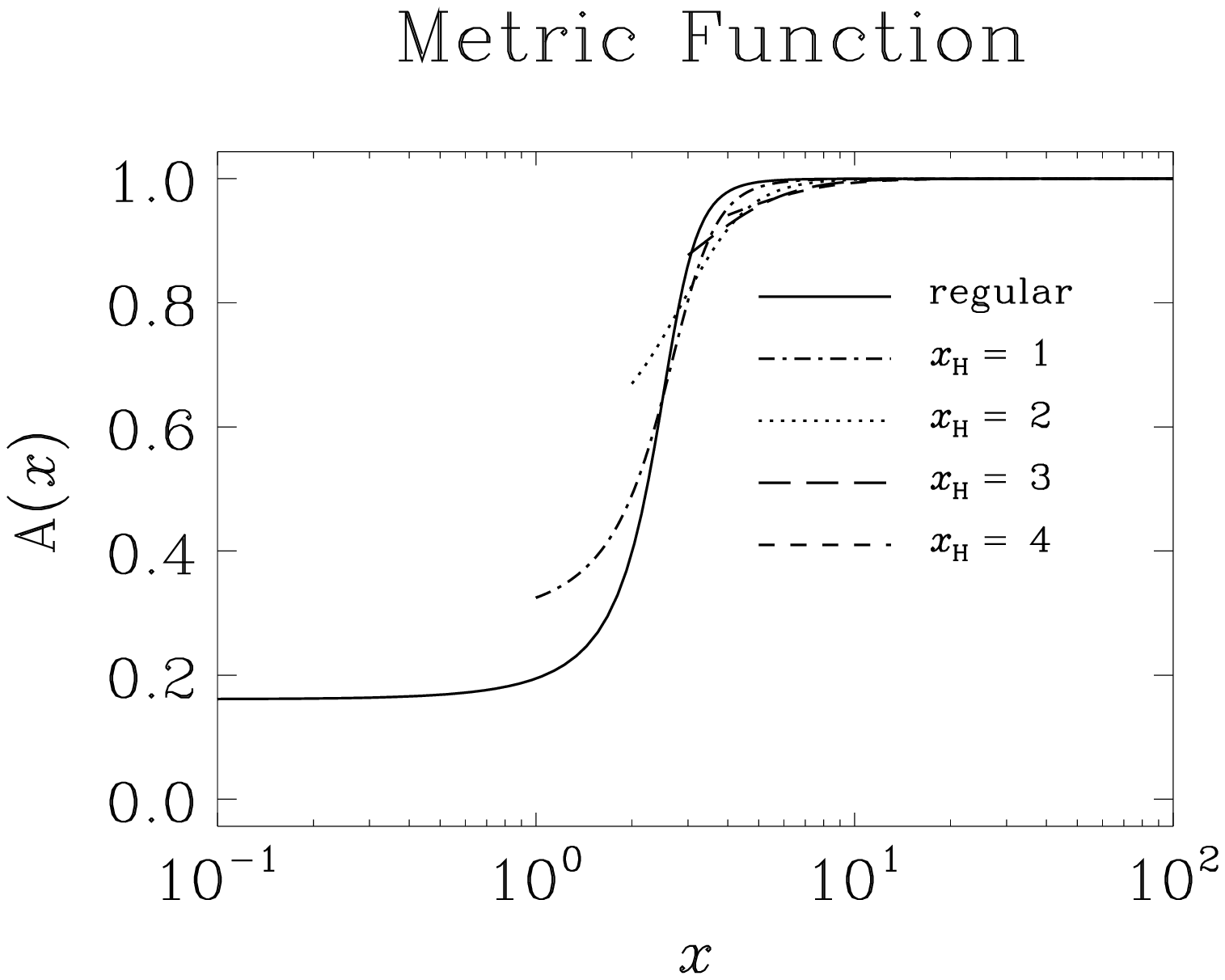}}
\caption{\label{a}
The function $A(x)$
is shown for the regular solution and for the
black hole solutions with horizons
$x_{\rm H}=1$, 2, 3 and 4
as a function of $x$.}
\end{figure}

In Figs.~\ref{k}-\ref{a} we show the lowest SO(3) solution.
It is obtained independently
in the limit of vanishing coupling constant
on the unstable upper branch of the Einstein-Skyrme system
\cite{kks}.
The excited solutions and
further details will be given elsewhere \cite{we}.

The instability of the regular solutions follows
from Theorem 1 of Ref.~\cite{strau3}.
There the instability of the solutions of K\"unzle \cite{kuenzle}
was demonstrated.
We find that the theorem applies also to the new class of
solutions with node structure $(0,n)$,
including the lowest mass solution
(where $\alpha=1$ \cite{strau3} as well).

\section{Black Hole Solutions}

We now turn to the black hole solutions of the
SU(3) Einstein-Yang-Mills system.
Imposing again the condition of asymptotic flatness,
the black hole solutions satisfy the same
boundary conditions at infinity
as the regular solutions.
The existence of a regular event horizon at $x_{\rm H}$
requires
\begin{equation}
\mu(x_{\rm H})= \frac{x_{\rm H}}{2}
\ , \end{equation}
and $A(x_{\rm H}) < \infty $,
and the matter functions must satisfy at the horizon $x_{\rm H}$
\begin{eqnarray}
 N'K'= \frac{1}{4 x^2} K \left( K^2+7H^2-4 \right)
\ , \\
 N'H'= \frac{1}{4 x^2} H \left( H^2+7K^2-4 \right)
\ . \end{eqnarray}

\begin{figure}
\centering
\epsfysize=11cm
\mbox{\epsffile{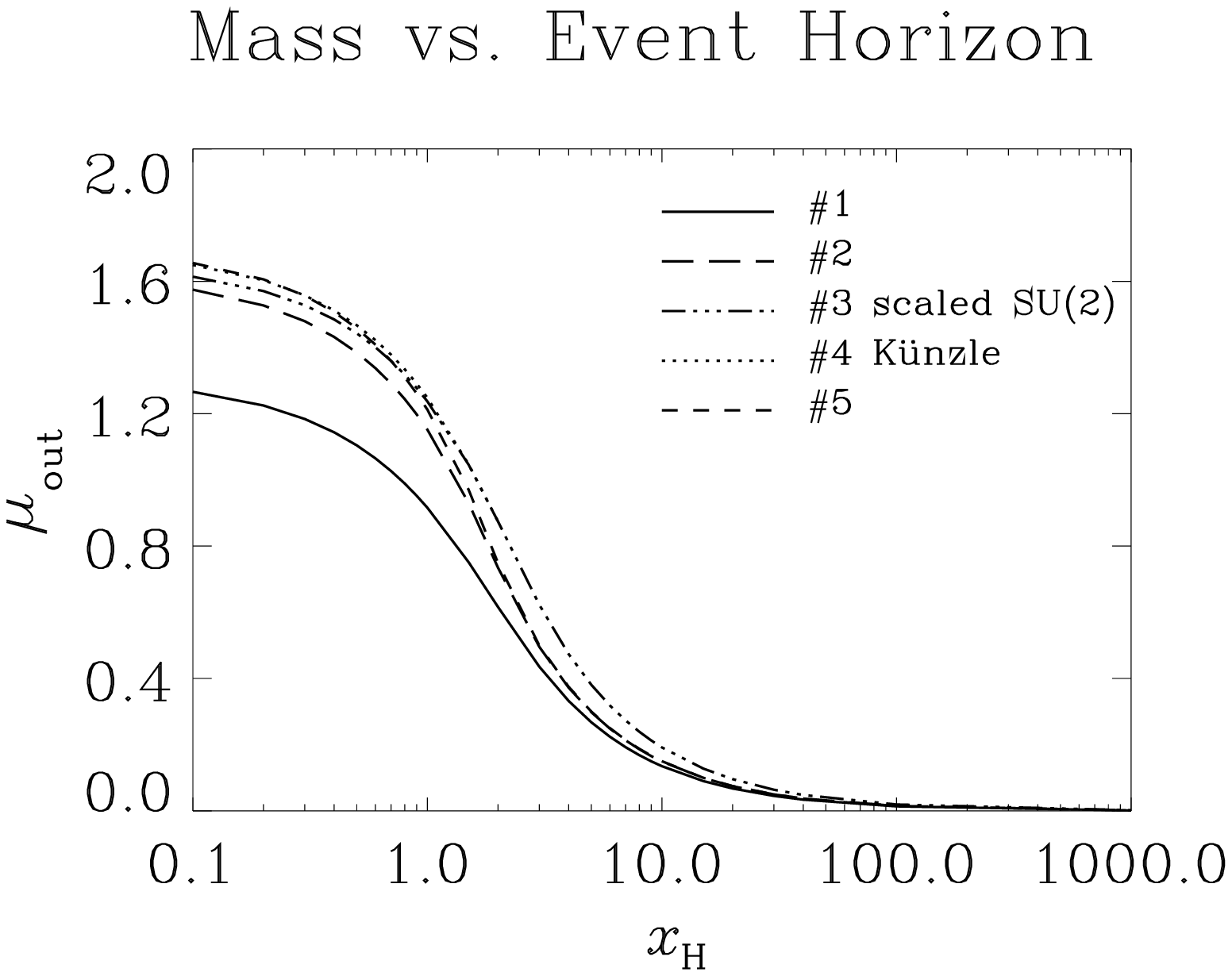}}
\caption{\label{engy}
The mass fraction outside the horizon, $\mu_{\rm out}$,
of the SO(3) black hole solutions of Table~2
is shown as a function of the horizon $x_{\rm H}$.
When the mass fraction within the horizon, $x_{\rm H}/2$, is added,
the ADM mass is obtained.}
\end{figure}

SO(3) black hole solutions have not been found previously
\cite{kuenzle}.
In Fig.~\ref{engy} we exhibit the masses of
the lowest SO(3) black holes in terms of the mass fractions outside
the horizon, $\mu_{\rm out}$, defined via
\begin{equation}
m_{\rm ADM} = \Bigl( \frac{x_{\rm H}}{2} + \mu_{\rm out} \Bigr)
  \frac{\sqrt{4 \pi}m_{\rm Pl}}{e}
= \mu(\infty) \frac{\sqrt{4 \pi}m_{\rm Pl}}{e}
\ , \end{equation}
as a function of the horizon $x_{\rm H}$.
For $x_{\rm H} \rightarrow 0$ the black hole solutions
approach the regular solutions.
With increasing horizon $x_{\rm H}$ these black hole solutions
keep their identity in terms of the boundary conditions
and the node structure.
Only the second solution with node structure $(1,1)$ (\#4 of Table~1)
disappears.
This solution has the same boundary conditions
and node structure as the lowest scaled SU(2) solution
(\#3 of Table~1), but a slightly higher mass.
It in fact merges into the scaled
SU(2) solution at a horizon of $x_{\rm H}=1.7146$,
leaving a unique black hole solution
with node structure $(1,1)$.
(The same feature holds for the two solutions
with node structure (2,2) of Ref.~\cite{kuenzle}.
The solution with higher mass merges into the solution
with lower mass, again a scaled SU(2) solution,
at a horizon of $x_{\rm H}=1.3745$.)
Note, that the order of the solutions changes
from the order of the regular solutions
as the horizon increases.
For instance, beyond $x_{\rm H}=0.703$ solution \#5
has a lower mass than solution \#3.

In Table~2 we present some properties of the black hole solutions
with a horizon $x_{\rm H}=1$,
emerging from the first five regular solutions of Table~1,
using again the notation of Ref.~\cite{kuenzle}.
The table contains the values of the functions $u_1$ and $u_2$
at the horizon, needed for a numerical shooting procedure.

The families of SO(3) black hole solutions change
continuously as a function of the horizon $x_{\rm H}$.
As examples we show the radial functions for the lowest
SO(3) black hole solutions
for the horizons $x_{\rm H}=1$, 2, 3, 4
in Figs.~(\ref{k})-(\ref{a}).
Further details of these solutions
and the excited SO(3) Einstein-Yang-Mills black holes
will be given elsewhere \cite{we}.

The instability of the black hole solutions follows
from Theorem 2 of Ref.~\cite{strau3}.
We find that the theorem applies to all black hole solutions
constructed
(where $\alpha=1$ \cite{strau3} as well).

\section{Conclusion}

The SU(3) Einstein-Yang-Mills system
possesses a sequence of regular
spherically symmetric solutions
based on the SO(3) embedding,
besides the well-studied sequence
based on the SU(2) embedding \cite{bm}.
The SO(3) solutions can be labelled
according to their node structure with two integers $(n_1,n_2)$
and the total number of nodes $n$.
The lowest solution has
node structure $(0,1)$ and $n=1$.
The first excited solution has node structure $(0,2)$,
while the second excited solution,
the lowest scaled SU(2) solution,
has node structure $(1,1)$.
The third excited solution \cite{kuenzle}
has the same node structure as the lowest scaled SU(2) solution,
being only slightly higher in mass.
The next solution then has $n=3$,
again with node structure $(0,n)$,
suggesting that this class of solutions has the lowest
mass for a given total number of nodes.

The regular SU(2) solutions are known to be unstable
\cite{strau1,volkov1},
the solution with $n$ nodes has $2n$ unstable modes \cite{lav,volkov4}.
The SO(3) solutions are unstable as well,
since Theorem 1 of Ref.\cite{strau3}
applies. It is an interesting open problem to study
the number of unstable modes
and find a relation to the number of nodes.

The lowest SU(2) solution has been interpreted
in analogy to the electroweak sphaleron \cite{km}
as the top of a barrier between vacua \cite{volkov1}.
Furthermore, like the electroweak sphaleron \cite{bk,kb},
the gravitating sphaleron also possesses
a fermion zero mode \cite{gib}
and gives rise to level-crossing \cite{gib,volkov2}.
It appears to be interesting to study fermions also
in the background of the lowest SO(3) solution.

Corresponding to each regular SO(3) solution there exist
black hole solutions.
These solutions keep their identity in terms of
the node structure, for arbitrary horizon.
If there are several solutions with the same structure
of nodes, solutions may disappear
by merging with the lowest solution of a given
node structure, as is for instance the case for
the lowest scaled SU(2) solution and its excitation
with node structure (1,1).

The SU(2) black holes are known to be unstable
\cite{strau2,volkov5,volkov4}.
The SO(3) black hole solutions are unstable as well,
since Theorem 2 of Ref.\cite{strau3}
applies.
It is an interesting open problem,
especially with respect to the bifurcations,
to study the number of unstable modes
of the SO(3) black holes.

Since the SU(3) Einstein-Yang-Mills system also contains
Schwarz\-schild black holes,
there are then many static, neutral black hole solutions
(including the SU(2) black holes) for a given mass,
enlarging the counterexample
to the ``no-hair conjecture''.
But only the Schwarz\-schild solution is stable.
The coloured black holes are all unstable.

Charged SU(3) black hole solutions
have been considered previously \cite{volkov3},
and SU(2)$\times$U(1) solutions have been constructed
\cite{volkov3}.
Here a natural extension is to consider
charged SO(3) black hole solutions.

{\sl Acknowledgement}

We gratefully acknowledge discussions
with M. Volkov.


\noindent
\begin{table}

\begin{tabular}{|c|cccclrr|c|} \hline
   &             & \multicolumn{2}{c}{nodes} &           &           &
   &               & \\ \cline{3-4}
   & $\tilde{\mu}(\infty)$ & $u_1$        & $u_2$      & $\beta_1$
& $\beta_2$ & $u_1(\infty)$ & $u_2(\infty)$ & \\ \hline
 1 & $0.65389$             & $0$          & $1$        & $-.41172$
& $0.25397$ & $1$           & $-1$          & \\
 2 & $0.81130$             & $0$          & $2$        & $-.60281$
& $0.39068$ & $1$           & $1$           & \\
 3 & $0.82865$             & $1$          & $1$        & $-.45372$
& $0$       & $-1$          & $-1$          & scaled SU(2) \\
 4 & $0.84769$             & $1$          & $1$        & $-.53766$
& $0.26077$ & $-1$          & $-1$          &  K\"unzle \\
 5 & $0.85237$             & $0$          & $3$        & $-.67272$
& $0.43953$ & $1$           & $-1$          & \\
 6 & $0.93774$             & $1$          & $2$        & $-.63437$
& $0.33228$ & $-1$          & $1$           & K\"unzle \\ \hline
\end{tabular}  \par
\vspace{0.5cm}

{\bf Table 1:} {
Properties of the lowest regular SO(3) solutions are given
in the notation of Ref.~\cite{kuenzle}.
The ADM mass is obtained from the second column via Eq.~(21) with
$\mu(\infty) = 2 \tilde \mu(\infty)$,
the third and forth column give the number of nodes
of the functions $u_1$ and $u_2$ defined in Eqs.~(22)-(23),
the fifth and sixth column provide the expansion coefficients
of these functions as defined in Eqs.~(24)-(25),
and the seventh and eighth column give the values
of the functions $u_1$ and $u_2$ at infinity.
}
\vspace{1.5cm}

\begin{tabular}{|c|cccclrr|c|} \hline
   &             & \multicolumn{2}{c}{nodes} &           &           &
   &               & \\ \cline{3-4}
   & $\tilde{\mu}(\infty)$ & $u_1$        & $u_2$
& $u_1(\tilde{x}_{\rm H})$ & $u_2(\tilde{x}_{\rm H})$
& $u_1(\infty)$ & $u_2(\infty)$ & \\ \hline
 1 & $0.70796$             & $0$          & $1$        & $0.92330$
& $0.90406$          & $1$           & $-1$          & \\
 2 & $0.82592$             & $0$          & $2$        & $0.88653$
& $0.84925$          & $1$           & $1$           & \\
 3 & $0.86855$             & $1$          & $1$        & $0.89723$
& $0.89723$          & $-1$          & $-1$          & scaled SU(2) \\
 4 & $0.87410$             & $1$          & $1$        & $0.89501$
& $0.87557$          & $-1$          & $-1$          & \\
 5 & $0.85608$             & $0$          & $3$        & $0.87481$
& $0.83035$          & $1$           & $-1$          & \\ \hline
\end{tabular}  \par
\vspace{0.5cm}

{\bf Table 2:} {
Properties of the lowest SO(3) black hole solutions are given
for the horizon $x_{\rm H}=1=2 \tilde x_{\rm H}$
following the classification of Table~1.
The ADM mass is obtained from the second column via Eq.~(21) with
$\mu(\infty) = 2 \tilde \mu(\infty)$,
the third and forth column give the number of nodes
of the functions $u_1$ and $u_2$ defined in Eqs.~(22)-(23),
the fifth and sixth column provide the value
of these functions at the horizon,
and the seventh and eighth column give the values
of the functions $u_1$ and $u_2$ at infinity.} \par
\end{table}

\newpage

\newpage

\newpage

\newpage

\newpage

\begin{thebibliography}{000}
\bibitem{bm}
 R. Bartnik, and J. McKinnon,
 Particlelike solutions of the Einstein-Yang-Mills equations,
 Phys. Rev. Lett. 61 (1988) 141.
\bibitem{strau1}
 N. Straumann, and Z.~H. Zhou,
 Instability of the Bartnik-McKinnon solutions
 of the Einstein-Yang-Mills equations,
 Phys. Lett. B237 (1990) 353.
\bibitem{volkov1}
 D.~V. Gal'tsov, and M.~S. Volkov,
 Sphalerons in Einstein-Yang-Mills theory,
 Phys. Lett. B273 (1991) 255.
\bibitem{lav}
 G. Lavrelashvili, and D. Maison,
 A remark on the instability of the Bartnik-McKinnon solutions,
 Phys. Lett. B343 (1995) 214.
\bibitem{volkov4}
 M.~S. Volkov, O. Brodbeck, G. Lavrelashvili, and N. Straumann,
 The number of sphaleron instabilities of the Bartnik-McKinnon solitons
 and nonabelian black holes,
 preprint ZU-TH-3-95, hep-th/9502045.
\bibitem{km}
 F.~R. Klinkhamer, and N.~S. Manton,
 A saddle-point solution in the Weinberg-Salam theory,
 Phys. Rev. D30 (1984) 2212.
\bibitem{volkov}
 M.~S. Volkov, and D.~V. Galt'sov,
 Black holes in Einstein-Yang-Mills theory,
 Sov. J. Nucl. Phys. 51 (1990) 747.
\bibitem{bizon1}
 P. Bizon,
 Colored black holes,
 Phys. Rev. Lett. 64 (1990) 2844.
\bibitem{kuenzle1}
 H.~P. K\"unzle and A.~K.~M. Masoud-ul-Alam,
 Spherically symmetric static SU(2) Einstein-Yang-Mills fields,
 J. Math. Phys. 31 (1990) 928
\bibitem{strau2}
 N. Straumann, and Z.~H. Zhou,
 Instability of colored black hole solutions,
 Phys. Lett. B243 (1990) 33.
\bibitem{volkov5}
 M.~S. Volkov, and D.~V. Gal'tsov,
 Odd-parity negative modes of Einstein-Yang-Mills
 black holes and sphalerons,
 Phys. Lett. B341 (1995) 279.
\bibitem{kuenzle}
 H.~P. K\"unzle,
 Analysis of the static spherically symmetric
 SU(n)-Einstein-Yang-Mills equations,
 Comm. Math. Phys. 162 (1994) 371.
\bibitem{kks}
 B. Kleihaus, J. Kunz, and A. Sood,
 SU(3) Einstein-Skyrme solitons and black holes,
 Utrecht preprint THU-95/6, hep-th/9503087.
\bibitem{bizon}
 P. Bizon, and T. Chmaj,
 Gravitating skyrmions,
 Phys. Lett. B297 (1992) 55.
\bibitem{strau3}
 O. Brodbeck, and N. Straumann,
 Instability proof for Einstein-Yang-Mills solitons
 and black holes with arbitrary gauge groups,
 ZU-TH-38-94, gr-qc/9411058.
\bibitem{we}
 B. Kleihaus, J. Kunz, and A. Sood, in preparation.
\bibitem{bk}
 J. Boguta, and J. Kunz,
 Hadroids and sphalerons,
 Phys. Lett. B154 (1985) 407.
\bibitem{kb}
 J. Kunz, and Y. Brihaye,
 Fermions in the background of the  sphaleron barrier,
 Phys. Lett. B304 (1993) 141.
\bibitem{gib}
 G.~W. Gibbons, and A.~R. Steif,
 Anomalous fermion production in gravitational collapse,
 Phys. Lett. B314 (1993) 13.
\bibitem{volkov2}
 M.~S. Volkov,
 Einstein-Yang-Mills sphalerons and level crossing,
 Phys. Lett. B334 (1994) 40.
\bibitem{volkov3}
 D.~V. Gal'tsov, and M.~S. Volkov,
 Charged non-abelian SU(3) Einstein-Yang-Mills black holes,
 Phys. Lett. B274 (1992) 173.

\end{thebibliography}
\end{document}